# Diffusion-assisted Model Predictive Control Optimization for Power System Real-Time Operation

Linna Xu
School of System Science and Engineering
Sun-Yat Sen University
Guangzhou, China
xuln6@mail2.sysu.edu.cn

Yongli Zhu
School of System Science and Engineering
Sun-Yat Sen University
Guangzhou, China
yzhu16@alum.utk.edu

*Abstract*—This paper presents a modified model predictive control (MPC) framework for real-time power system operation. The framework incorporates a diffusion model tailored for time series generation to enhance the accuracy of the load forecasting module used in the system operation. In the absence of explicit state transition law, a model-identification procedure is leveraged to derive the system dynamics, thereby eliminating a barrier when applying MPC to a renewables-dominated power system. Case study results on an industry park system and the IEEE 30-bus system demonstrate that using the diffusion model to augment the training dataset significantly improves load-forecasting accuracy, and the inferred system dynamics are applicable to the real-time grid operation with solar and wind.

## I. Introduction

As the scale and complexity of power systems continue to increase, ensuring their safe, stable, and efficient operation has become a key challenge. In recent years, model predictive control (MPC) [1] has become a prevailing tool for real-time dispatch and optimization of power systems due to its ability to tackle varying constraints caused by system evolutions. For example, in [2], the author employs the MPC in a decentralized voltage control framework for the coordination of the PV (photovoltaic) and EV (electric vehicle). In [3], a distributed MPC approach is adopted for real-time voltage regulation in EV-penetrated distribution networks. In [4], MPC is applied in highway transportation power systems to optimize its voltage and power flow at a time scale of seconds level.

Though the MPC method has proven itself in various applications, it requires known and explicit system dynamics, i.e., the state transition matrix. However, the system transition matrix is not readily available in real-world power systems, especially for renewable sources (e.g., solar and wind) with inherent stochasticity and volatility. Hence, the applicability of the MPC framework will be compromised for renewable sources with unknown state transition laws.

Another concern for real-time system operation is short-term load forecasting. To achieve more reliable operational effects, a forecasting model with higher accuracy is expected. However, high-quality measurement data can be scarce, making it hard to train a highly performant load forecaster.

Therefore, proposing a more flexible MPC framework for more efficient and reliable system operation is of great value. To that end, this paper proposes an MPC optimization framework based on state-space model identification and TS-Diffusion [5] to enhance the applicability of MPC for the system's real-time operation and load forecasting accuracy. The diffusion model is mainly used to augment a limited time series dataset in order to train an accurate short-term load forecaster for system real-time operation.

In the second section, the basic idea of the TS-Diffusion model used for load-forecasting data augmentation is explained. The third section introduces the state-space model identification approach to reconstruct the unknown state transition law for the MPC framework. Then, case studies are carried out on an industry park system and the IEEE 30-bus system. The results of model identification, load forecasting, and system power dispatch are presented in the fourth section. The final section concludes the whole paper with future work envisioned.

## II. Diffusion Model for Load Prediction

### A. Principle of TS-Diffusion

The diffusion model is a novel generative artificial intelligence methodology that has emerged in recent years [6][7]. TS-Diffusion is one kind of diffusion model designed for time series data generation, leveraging a Transformer-inspired architecture combined with a decomposition approach. It demonstrates strong performance in missing-value interpolation, making it suitable for time series augmentation. The training process of the TS-Diffusion model is illustrated in Fig.1.

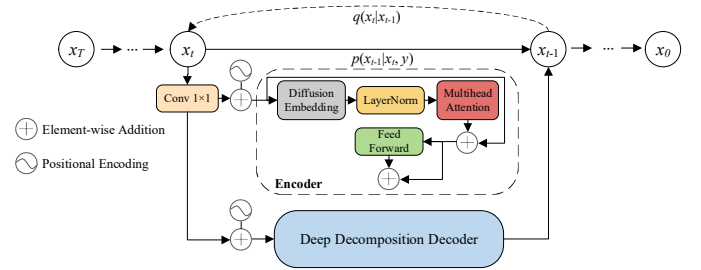

Fig. 1. The structure diagram of TS-Diffusion.

As shown in Fig.1, the TS-Diffusion model contains forward and reserve processes. The encoder module processes the input time series through a multi-head attention mechanism combined with a feed-forward neural network. Similarly, the decoder module utilizes multi-head attention and feed-forward layers, augmented by a deep decomposition design that isolates trend and seasonality components of the time series. The diffusion-embedding module integrates time-step information, while the positional encoding module adds spatial context, enabling the model to capture the latent temporal patterns within the data effectively.

In this framework, a sample drawn from the data distribution $x_0 \sim q(x)$ is incrementally corrupted with Gaussian noise $\mathcal{N}$ over a forward diffusion process. At each step $t$, the transition is defined by $q(x_t|x_{t-1}) = \mathcal{N}(x_t; \sqrt{1-\beta_t} x_{t-1}, \beta_t I)$, where $\beta_t \in (0,1)$ specifies the level of noise introduced at each step. Subsequently, a neural network is trained to model the reverse process, learning to gradually denoise the sample by approximating the reverse transition $q(x_t|x_{t-1}) = \mathcal{N}(x_t; \mu_\theta(x_t,t), \Sigma_\theta(x_t,t))$. This reverse process of progressively removing noise can be effectively approximated via (1).

$$x_{t-1} = \frac{\sqrt{\bar{\alpha}_{t-1}} \beta_t}{1-\bar{\alpha}_t} \hat{x}_0(x_t,t,\theta) + \frac{\sqrt{\alpha_t}(1-\bar{\alpha}_{t-1})}{1-\bar{\alpha}_t} x_t + \frac{1-\bar{\alpha}_{t-1}}{1-\bar{\alpha}_t} \beta_t z_t \quad (1)$$

where $z_t \sim \mathcal{N}(0, I)$, $\alpha_t = 1-\beta_t$ and $\bar{\alpha}_t = \prod_{s=1}^{t} \alpha_s$. TS-Diffusion trained the denoising model $\mu_\theta(x_t, t)$ using a loss function of weighted mean-squared-error, as described in (2).

$$\mathcal{L}_{simple} = \mathbb{E}_{t,x_0} \left[ w_t \| x_0 - \hat{x}_0(x_t,t,\theta) \|^2 \right], w_t = \frac{\lambda \alpha_t (1-\bar{\alpha}_t)}{\beta_t^2} \quad (2)$$

where $\lambda$ is a constant. These loss terms are down-weighted at small $t$ to force the network focus on a larger diffusion step. In addition, TS-Diffusion guides an interpretable diffusion training by applying the Fourier transformation in the frequency domain, i.e.,

$$\mathcal{L}_\theta = \mathbb{E}_{t,x_0} \left[ w_t \cdot \begin{pmatrix} \lambda_1 \| x_0 - \hat{x}_0(x_t,t,\theta) \|^2 + \\ \lambda_2 \| \mathcal{FFT}(x_0) - \mathcal{FFT}(\hat{x}_0(x_t,t,\theta)) \|^2 \end{pmatrix} \right] \quad (3)$$

where $\mathcal{FFT}$ denotes the Fast Fourier Transformation, and $\lambda_1$, $\lambda_2$ are the balancing weights for the two losses in (3).

### B. Framework of Load Forecasting based on TS-Diffusion

In this paper, we use the ExtraTree [8] model to predict load demand and comprehensively evaluate the effectiveness of the augmented dataset by TS-Diffusion. The overall framework for dataset splitting and augmentations is illustrated in Fig.2.

As shown in Fig. 2, the original load dataset is initially divided into training and testing sets. The training set is then expanded by applying the TS-Diffusion method, which aims to enhance the diversity and representativeness of the given dataset.

The load forecasting model is then trained by combining the original training set and the TS-Diffusion-generated data. The testing set remains untouched throughout the generating and training process, serving as a rigorous benchmark to objectively evaluate the model's true performance. Finally, prediction metrics are calculated to observe the improvements in the predictive accuracy gained from the generated data.

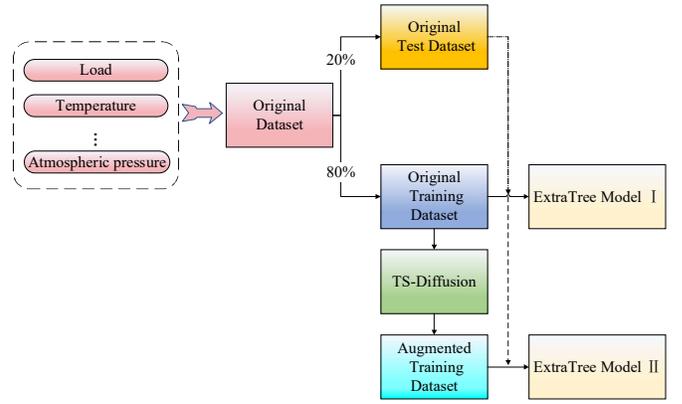

Fig. 2. Illustration of dataset splitting and augmentation.

### III. MPC OPTIMIZATION BASED ON INFERRED STATE TRANSITION LAW

#### A. Dataset Description

The dataset used in this paper is sourced from an industry park that receives its energy from a combination of grid-supplied electricity and on-site wind power generation. This mixed energy supply provides a practical setting for analyzing hybrid renewable power systems. A schematic diagram illustrating the energy flow of this park is depicted in Fig. 3.

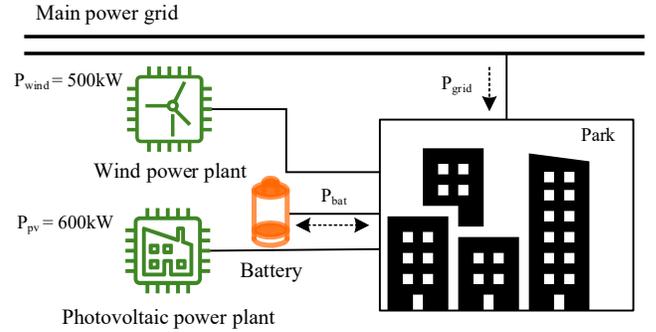

Fig. 3. The schematic diagram of a renewable-supplied industry park.

The data is collected at an hourly resolution, capturing the load demand and the upper limit of wind power generation over a typical week (168 hours) for the park. These data points serve as the basis for training the load diffusion and wind power diffusion models, respectively, allowing for a detailed analysis of both load and renewable generation dynamics.

#### B. Problem Modeling

The MPC-based optimization model established in this paper is as follows:

$$\min_{P_{pv}, P_{wind}, P_{grid}, P_{bat}} \sum_{t=k}^{k+N-1} c_{grid} P_{grid}(t) + c_{pv} P_{pv}(t) + c_{wind} P_{wind}(t) + c_{bat} |P_{bat}(t)| \quad (4)$$

$$s.t. \ P_{load}(t) = P_{grid}(t) + P_{pv}(t) + P_{wind}(t) + P_{bat}(t), t=k,...,k+N-1 \quad (5)$$

$$\begin{cases} P_{bat}(t) \geq 0, P_{discharge}(t) = P_{bat}(t), P_{charge}(t) = 0 \\ P_{bat}(t) \leq 0, P_{charge}(t) = -P_{bat}(t), P_{discharge}(t) = 0 \end{cases}, t=k,...,k+N-1 \quad (6)$$

$$SOC(t) = SOC(t-1) + \eta_{cha} \frac{P_{charge}(t)}{E_{max}} - \frac{P_{discharge}(t)}{\eta_{dis} E_{max}}, t=k,...,k+N-1 \quad (7)$$

$$P_{grid}(t) \geq 0, t = k,...,k+N-1 \quad (8)$$

$$0 \leq SOC(t) \leq 1, t = k,...,k+N-1 \quad (9)$$

$$-P_{bat,max} \leq P_{bat}(t) \leq P_{bat,max}, t = k,...,k+N-1 \quad (10)$$

$$0 \leq P_{pv}(t) \leq P_{pv,max}(t), t = k,...,k+N-1 \quad (11)$$

$$0 \leq P_{wind}(t) \leq P_{wind,max}(t), t = k,...,k+N-1 \quad (12)$$

$$\begin{bmatrix} P_{pv,max}(t) \\ P_{wind,max}(t) \end{bmatrix} = A \begin{bmatrix} P_{pv,max}(t-1) \\ P_{wind,max}(t-1) \end{bmatrix} + B \begin{bmatrix} P_{pv}(t-1) \\ P_{wind}(t-1) \end{bmatrix}, t = k,...,k+N-1 \quad (13)$$

The objective function in (4) represents a *fictitious* total generation cost of the park over $N$ hours ($N$=8 in this paper) in the *fictitious* future. This cost comprises four components: the cost of purchasing electricity from the main grid, the cost of wind power generation, the cost of photovoltaic power generation, and the cost of energy storage operation. $c_{grid}$ is set to 1\$/kWh, $c_{pv}$ is set to 0.4\$/kWh, $c_{wind}$ is set to 0.5\$/kWh, and $c_{bat}$ is set to 1\$/kWh.

In this MPC framework, the state variables are defined as the upper limits of wind and photovoltaic power at each time step $t$, denoted as $P_{wind,max}(t)$, $P_{pv,max}(t)$. The control variables are the actual wind power output $P_{wind}(t)$ and the actual photovoltaic power output $P_{pv}(t)$. This MPC framework solves a series of sub-problems of the above form by iterating over $k$ until a final time index (e.g., for day-ahead, hourly resolution, $k$ runs from 1 to 24). At the end of each sub-problem, the *first* value of the decision variable is recorded and utilized in the next period. The *genuine* objective value (cf. cost items in (4)) is *finally* constructed based on the recorded decision variables during each intermediate period.

The time-dependent transition logic for the battery storage's $SOC$ is expressed in (7), where $\eta_{cha}$ and $\eta_{dis}$ are respectively the charging and discharging efficiencies. Considering the inherent characteristics of the battery, $\eta_{cha}$ is typically lower than $\eta_{dis}$. Thus, $\eta_{cha}$ is set to 0.95, and $\eta_{dis}$ is set to 0.99 in this paper. $E_{max}$ is the battery's energy capacity and is set to 100kWh. $P_{bat,max}$ is the battery's maximum power and is set to 50kW.

In (13), $A$ and $B$ represent the state transfer matrix and control matrix, which are inferred from the historical wind and photovoltaic power data via a model-identification procedure, viz., *iterative rational function estimation* and *prediction error minimization* from the system identification theory [9], as described in (14) to (16). More specifically, for a given single-input-single-output system in (14), the model identification procedure is briefly described in (15) (16).

$$\begin{aligned} x(k+1) &= Ax(k) + Bu(k) \\ y(k) &= Cx(k) + Du(k) \Leftrightarrow Gu(k) + He(k) \end{aligned} \quad (14)$$

$$G(q) = C(qI_{nx} - A)^{-1}B + D, H(q) = C(qI_{nx} - A)^{-1}K + I_{ny} \quad (15)$$

$$V_N(G,H) = \sum_{i=1}^{N} e^2(t), e(t) = H^{-1}(q)(y(t) - G(q)u(t)) \quad (16)$$

where, $e(t)$ is the difference between the system model's measured output and predicted output. The error for a linear model is defined as (17), where $V_N(G,H)$ is a cost function. The subscript $N$ indicates the number of data samples. $H$ and $G$ are, respectively, the transfer function relating the control input $u$ and the error $e$ to the output $y$. By minimizing $V_N(G,H)$, the optimal $G$ and $H$ can be found, and finally, the matrices $A$ and $B$ can be determined. $nx$ and $ny$ are, respectively, the number of states and outputs. $I_{nx}$ and $I_{ny}$ are identity matrices with proper dimensions.

The overall MPC framework, constructed based on the estimated state transition matrix and control matrix, is illustrated in Fig.4. As shown in Fig.4, we need an adequate set of high-quality historical data, i.e., a consecutive series of $[x(t), x(t-1)]$ pairs to infer the state matrix and control matrix, where $x(t):=[P_{pv,max}(t), P_{wind,max}(t)]^T \in \mathbb{R}^2$. In this paper, the following way is adopted: from the initially given dataset, recursively feeding the previous-day values (i.e., $P_{wind,max}(t)$ and $P_{pv,max}(t)$) and solving a **non-MPC** optimization problem (i.e., without (13)) to obtain a series of $P_{wind}(t)$ and $P_{pv}(t)$. In this way, the historical time series needed to identify matrices $A$ and $B$ can be obtained.

After obtaining matrices $A$ and $B$, (13) is added to the solving process of the subsequent MPC sub-problems. The $A$, $B$ matrices can be fixed (i.e., inferred just one time at the very beginning of the optimization horizon) or *dynamically inferred* in a rolling manner: after the optimization of a previous period (say, $t$-$N$~$t$-1) is finished, the corresponding state and control variables are utilized to *re-infer* $A$ and $B$ for the next period (i.e., $t$~$t$+$N$-1), and the similar process repeats until the final period.

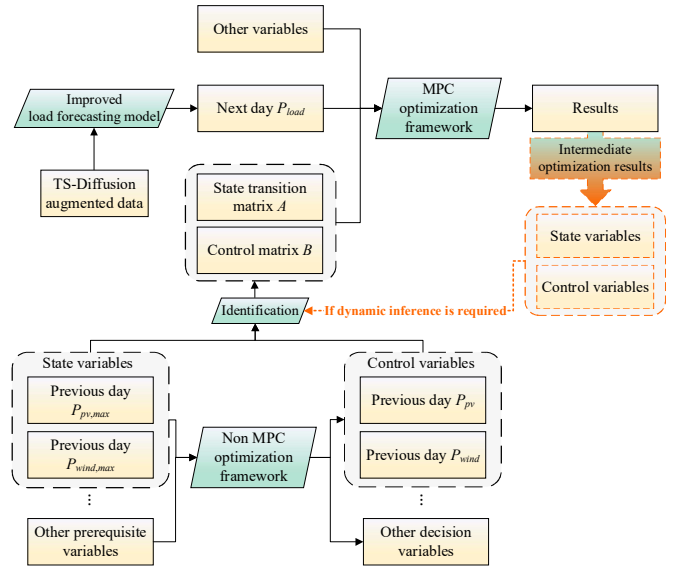

Fig.4. The flowchart of the proposed MPC framework for system operation.

## IV. EXPERIMENTS AND RESULTS

### A. Experiment Setup

In the experiment here, the initial energy storage $SOC$ is set to 50%, the photovoltaic power generation capacity is set to 600kW, and the wind power generation capacity is set to 500kW. All optimization models are implemented in MATLAB 2023b.

The TS-diffusion and load forecasting models' training process and data generation are conducted in PyTorch on a desktop PC with an Intel 5.4GHz CPU, 32GB RAM, and 4060ti GPU. The power system optimization model is executed on a laptop with an AMD Ryzen 7 8845HS CPU and 16GB RAM.

## B. Results of TS-Diffusion-assisted Load Forecasting

To inspect the potential impact of overfitting due to the increase in data size, we conducted an additional set of experiments. In these experiments, the original training set was artificially expanded by duplicating itself to match the size of the training set expanded by the TS-Diffusion model. This approach provides a fair comparison between the original and TS-Diffusion-augmented datasets [10]. The hyperparameters for TS-diffusion model training are listed in Table I. Under such settings, the total time required to train the TS-Diffusion model and then augment the dataset is about 40 minutes (done offline *before* running the optimization model).

TABLE I. HYPERPARAMETERS FOR TRAINING THE TS-DIFFUSION MODEL

| Hyperparameter | Value |
| --- | --- |
| Encoder Layer | 4 |
| Decoder Layer | 3 |
| Kernel Size | 1 |
| Learning Rate | 1.0e-5 |
| Solver Gradient Accumulation | 2 |
| Scheduler Threshold | 1.0e-1 |

Table II presents a detailed comparison of the error metrics for the load forecasting model (ExtraTree) trained on different training sets, highlighting the effect of data augmentation on the model's forecasting capability.

TABLE II. COMPARISON RESULTS OF DIFFERENT TRAINING DATASETS APPLIED TO THE LOAD FORECASTING MODEL

| Dataset | MAE | RMSE |
| --- | --- | --- |
| original | 0.04467 | 0.03209 |
| replicated | 0.04495 | 0.03229 |
| augmented | **0.00023** | **0.00004** |

As shown in Table II, the dataset augmented using the TS-Diffusion yields much better performance than the original dataset in terms of lower MAE (mean absolute error) and RMSE (root mean squared error). Additionally, the performance of the TS-Diffusion-augmented dataset surpasses that of the dataset expanded by simple replication. This result reveals that the improvement in the model's predictive accuracy is not merely due to an increase in the training set size but is attributed to the enhanced quality of the data generated by the TS-Diffusion. Finally, the predicted load curves on the testing set obtained using different training sets are shown in Fig. 5.

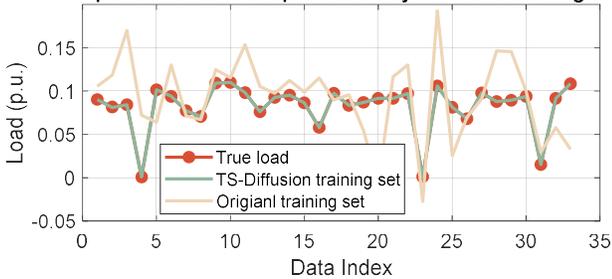

Fig.5. Prediction comparison on the testing set based on different training sets.

In Fig. 5, the red circle represents the actual load value in the testing set, the green line represents the performance of the forecasting model trained with the TS-Diffusion augmented dataset on the testing set, and the yellow line represents the forecasted curves using the original dataset. It can be seen that curve fitness by training with the original dataset is not as good as the TS-Diffusion augmented dataset. The green line matches the red circle better at every index in the figure, consistent with the error comparison results in Table I.

## C. Results on the Industry-Park System

When the next day's photovoltaic, wind, and load demand conditions are not known in advance, we can utilize the load forecasting model trained on the TS-Diffusion augmented data (in subsection B) and the inferred state transition matrices to establish the MPC framework. Fig. 6 (top) displays the 24-hour power dispatch based on the proposed MPC framework and the TS-Diffusion-assisted load forecasting model. The total cost is 4847.1$. When the system model is dynamically inferred (every 8 hours) in the proposed MPC framework, the results are shown in Fig 6 (middle), and the total cost is 4801.4$.

To verify the effectiveness of the proposed methods, we also set up a benchmark using a non-MPC optimization framework (i.e., without (13) and fixed $P_{pv,max}$ and $P_{wind,max}$ as their initial values) to solve the problem under the same load demand. The results are shown in Fig. 6 (bottom). The total cost is 4905.2$.

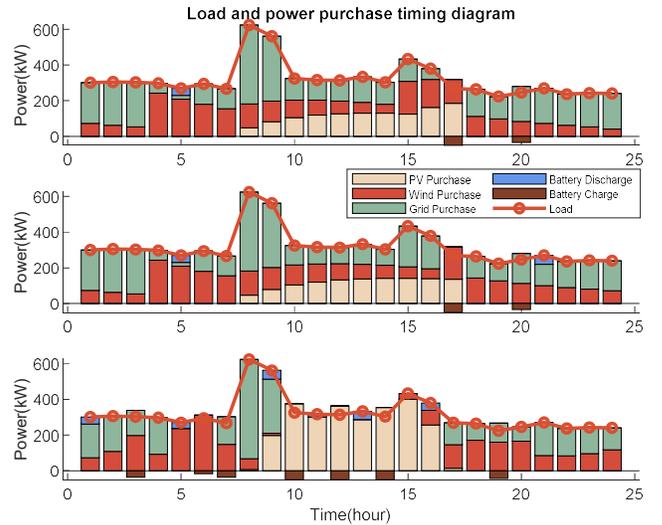

Fig. 6. The power dispatch diagram obtained respectively by: the proposed MPC, the proposed MPC with dynamic inference, a non-MPC framework.

Table III lists all the objective values (total generation cost) obtained by different approaches under the same load demand.

TABLE III. OBJECTIVE VALUES OBTAINED BY DIFFERENT APPROACHES

| Methods | Objective Value |
| --- | --- |
| Benchmark | 4905.2$ |
| Proposed MPC | 4847.1$ |
| Proposed MPC with dynamic inference | 4801.4$ |

## D. Results on the IEEE 30-bus System

In order to validate the performance of our proposed approach on a more complicated system, we conduct corresponding experiments on the IEEE 30-bus system. The load demand curve obtained by the previous TS-Diffusion-assisted load forecasting model is still used here, but its scale of magnitude is adjusted to match the original load amount of the IEEE-30 system.

The generators connected to bus-2 and bus-23 are modeled as wind and photovoltaic generators, respectively. Based on this setting, we then solve the optimization model (4) - (13) with the objective function modified as (17), and add network topology-related constraints (19)-(21) as follows (other constraints are similar to section II.B, hence omitted here):

$$\min_{P_g, P_{wind}, P_{pv}, \delta} \sum_{t=k}^{k+N-1} c_{grid} \cdot P_g(i,t) + c_{pv} \cdot P_{pv}(t) + c_{wind} \cdot P_{wind}(t) + c_{bat} \cdot P_{bat}(t) \quad (17)$$

$$\text{s.t.} \quad P_{g\min}(i,t) \leq P_g(i,t) \leq P_{g\max}(i,t) \quad (18)$$

$$P_{ij}(t) = B_{ij}(\delta_i(t) - \delta_j(t)) \quad (19)$$

$$P_{ij\min}(t) \leq P_{ij}(t) \leq P_{ij\max}(t) \quad (20)$$

$$P_g(i,t) - P_d(i,t) = \sum_j B_{ij} \Delta \delta_{ij}(t) \quad (21)$$

In (17), $P_g(i,t)$ is the power generation of the $i$-th non-renewable generator at time $t$. In (19), $P_{ij}(t)$ is the power flow from bus $i$ to bus $j$ at time $t$, $\delta_i(t)$ is the phase angle of bus $i$ at time $t$, and $B_{ij}$ is the line susceptance between bus $i$ and bus $j$. In (21), $P_d(i,t)$ means the load power of bus $i$ at time $t$. Energy storage is added to bus-3.

Fig. 7 (top) displays the 24-hour power dispatch based on the proposed MPC framework and the TS-Diffusion-assisted load forecasting model. The total cost is 5196.4$. When the system model is dynamically inferred (every 8 hours) in the proposed MPC framework, the results are shown in Fig 7 (middle), and the total cost is 5196.3$.

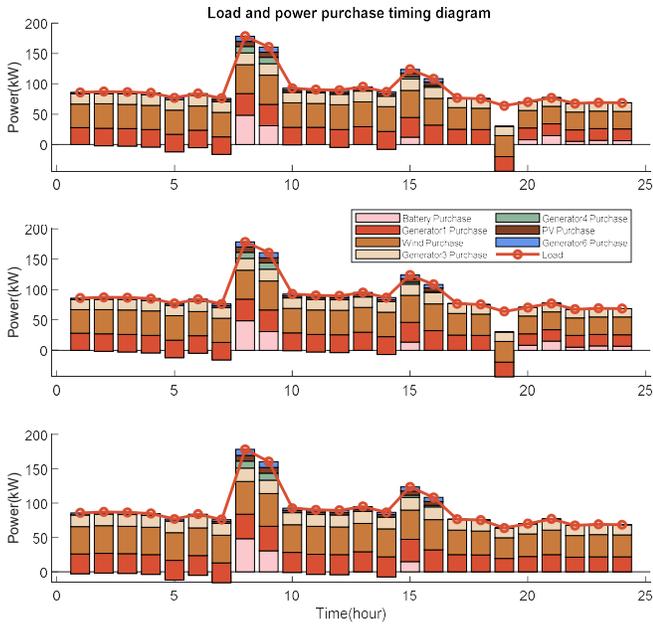

Fig. 7. The power dispatch diagram for the IEEE 30-bus system obtained respectively by: the proposed MPC, the proposed MPC with dynamic inference, a non-MPC framework.

Similar to the previous case study, the power dispatch results of the benchmark (using a non-MPC optimization framework to solve the problem under the same load demand) are shown in Fig. 7 (bottom). The total cost is 5287.4$.

Table IV lists the objective values (total generation cost) obtained by different approaches.

TABLE IV. OBJECTIVE VALUES OBTAINED BY DIFFERENT APPROACHES ON THE IEEE 30-BUS SYSTEM

| Methods | Objective Value |
|---|---|
| Benchmark | 5287.4$ |
| Proposed MPC | 5196.4$ |
| Proposed MPC with dynamic inference | 5196.3$ |

## V. CONCLUSION

This paper introduces a new MPC-based optimization framework that addresses the challenge of unknown state transition equations in traditional MPC setups by reconstructing the state transition matrices on given time series data. Additionally, integrating the TS-Diffusion method enhances the accuracy of the load forecasting model, contributing to the credibility of the optimization outcome when used for system real-time operation. The next step is to experiment with other data-augmentation approaches and investigate nonlinear system models of renewable sources in the proposed MPC framework.

REFERENCES

[1] D. Erazo-Caicedo, E. Mojica-Nava, and J. Revelo-Fuelagán, "Model predictive control for optimal power flow in grid-connected unbalanced microgrids," *Electr. Power Syst. Res.*, vol. 209, p. 108000, 2022.

[2] L. Wang, A. Dubey, A. H. Gebremedhin, A. K. Srivastava, and N. Schulz, "MPC-Based Decentralized Voltage Control in Power Distribution Systems With EV and PV Coordination," *IEEE Trans. Smart Grid*, vol. 13, no. 4, pp. 2908–2919, 2022, doi: 10.1109/TSG.2022.3156115.

[3] J. Hu, C. Ye, Y. Ding, J. Tang, and S. Liu, "A Distributed MPC to Exploit Reactive Power V2G for Real-Time Voltage Regulation in Distribution Networks," *IEEE Trans. Smart Grid*, vol. 13, no. 1, pp. 576–588, 2022.

[4] W. Huang, C. Gao, R. Li, R. Bhakar, N. Tai, and M. Yu, "A Model Predictive Control-Based Voltage Optimization Method for Highway Transportation Power Supply Networks With Soft Open Points," *IEEE Trans. Ind. Appl.*, vol. 60, no. 1, pp. 1141–1150, 2024.

[5] X. Yuan and Y. Qiao, "Diffusion-TS: Interpretable Diffusion for General Time Series Generation," *Int. Conf. Learn. Represent. (ICLR)*, 2024. [Online]. Available: https://openreview.net/forum?id=4h1apFjO99.

[6] J. Ho, A. Jain, and P. Abbeel, "Denoising diffusion probabilistic models," The 34th Conference on Neural Information Processing Systems (NeurIPS 2020), Vancouver, Canada, 2020. [Online]. Available: https://doi.org/10.48550/arXiv.2006.11239.

[7] Y. Song, J. Sohl-Dickstein, D. P. Kingma, A. Kumar, S. Ermon, and B. Poole, "Score-Based Generative Modeling through Stochastic Differential Equations," *Int. Conf. Learn. Represent. (ICLR)*, 2021.

[8] P. Geurts, D. Ernst, and L. Wehenkel, "Extremely randomized trees," *Mach. Learn.*, vol. 63, no. 1, pp. 3–42, 2006.

[9] L. Ljung, System Identification: Theory for the User, Second Edition. Upper Saddle River, NJ: Prentice Hall PTR, 1999.

[10] L. Xu, Y. Zhu. Generative Modeling and Data Augmentation for Power System Production Simulation. NeurIPS 2024 D3S3 workshop. https://neurips.cc/virtual/2024/105168.